\documentclass[aps,pra,onecolumn,superscriptaddress,longbibliography,nofootinbib,showkeys,showpacs]{revtex4-1}
\usepackage{amsfonts}
\usepackage{amsmath}
\usepackage{appendix}
\usepackage{bm}
\usepackage{color}
\usepackage{braket}
\usepackage{url}
\usepackage{xcolor}
\usepackage{comment}
\usepackage{graphicx}
\usepackage{hyperref}
\usepackage[utf8]{inputenc}
\usepackage{mathptmx}
\usepackage{CJK}
\usepackage[capitalise]{cleveref}
\usepackage{subfigure}
\usepackage{cases}
\usepackage[ruled,vlined,linesnumbered,boxed]{algorithm2e}

\begin{document}
\colorlet{CV}{.}
\SetKwRepeat{Do}{do}{while}
\title{Ising Hamiltonians for Constrained Combinatorial Optimization Problems and the Metropolis-Hastings Warm-Starting Algorithm}

\author{Hui-Min Li}
\affiliation{School of Mathematical Sciences, Capital Normal University, 100048, Beijing, China}
\author{Jin-Min Liang}
\affiliation{School of Mathematical Sciences, Capital Normal University, 100048, Beijing, China}
\author{Zhi-Xi Wang}
\email{wangzhx@cnu.edu.cn}
\affiliation{School of Mathematical Sciences, Capital Normal University, 100048, Beijing, China}
\author{Shao-Ming Fei}
\email{feishm@cnu.edu.cn}
\affiliation{School of Mathematical Sciences, Capital Normal University, 100048, Beijing, China}

\begin{abstract}
Quantum approximate optimization algorithm (QAOA) is a promising
variational quantum algorithm for combinatorial optimization problems. However, the implementation of QAOA is limited due to the requirement that the problems be mapped to Ising Hamiltonians and the nonconvex optimization landscapes. Although the Ising Hamiltonians for many NP hard problems have been obtained, a general method to obtain the Ising Hamiltonians for constrained combinatorial optimization problems (CCOPs) has not yet been investigated. In this paper, a general method is introduced to obtain the Ising Hamiltonians for CCOPs and the Metropolis-Hastings warm-starting algorithm for QAOA is presented which can provably converge to the global optimal solutions. The effectiveness of this method is demonstrated by tackling the minimum weight vertex cover (MWVC) problem, the minimum vertex cover (MVC) problem, and the maximal independent set problem as examples. The Ising Hamiltonian for the MWVC problem is obtained first time by using this method. The advantages of the Metropolis-Hastings warm-starting algorithm presented here is numerically analyzed through solving 30 randomly generated MVC cases with 1-depth QAOA.

\end{abstract}

\maketitle

\section{Introduction}

Recently, there is immense interest in solving combinatorial optimization problems by using quantum algorithms, such as quantum adiabatic algorithms \cite{RevModPhys.80.1061,Farhi2001AQA,mehta_quantum_2021}, digitized adiabatic quantum computation \cite{Barends2016DigitizedAQ,chandarana_digitized-counterdiabatic_2022}, quantum approximate optimization algorithm (QAOA) \cite{farhi_quantum_2014,wurtz_maxcut_2021,crooks_performance_2018}, and variational quantum eigensolver (VQE) \cite{peruzzo_variational_2014,cerezo_variational_2021}. However, these quantum algorithms usually require that the classical combinatorial optimization problems be mapped to Ising Hamiltonians whose ground states encode the solutions to these problems. Although the Ising Hamiltonians for many NP hard problems have been obtained in \cite{lucas_ising_2014}, a general method to obtain the Ising Hamiltonians for constrained combinatorial optimization problems (CCOPs) has not yet been investigated.

In particular, the QAOA was originally proposed to find the approximate solutions to combinatorial optimization problems by Farhi $\emph{et al.}$ in 2014 \cite{farhi_quantum_2014}. It is a promising approach for near-term noisy intermediate-scale quantum (NISQ) devices \cite{Preskill2018quantumcomputingin} which are lack of error corrections and have imperfect gate implementations. In QAOA, the approximate solution to a combinatorial optimization problem can be encoded as an ansatz wave function with $2p$ angle parameters ($p$ is the depth of the algorithm). This wave function can encode the exact solution when $p\to\infty$ according to the adiabatic theorem \cite{farhi_quantum_2000}. Since each gate operation involves a certain amount of noise and QAOA is an hybrid quantum-classical algorithm which requires an outer loop classical optimization to assign angle parameters, it becomes more challenging to apply QAOA with high-depth. To reduce the circuit depth in QAOA is of importance and interesting. There are different methods for such proposal, such as introducing  digitized counterdiabatic technique \cite{hegade_digitized_2022,hegade_portfolio_2022} into QAOA.

Here, we focus on improving the performance of original QAOA with low-depth. While low-depth QAOA may achieve promising results for some problems \cite{wurtz_maxcut_2021}, the optimization landscapes would become more nonconvex so that it tends to obtain the local optimal solutions which plagues QAOA \cite{PhysRevA.104.032401,patti_markov_2021}. Recently, the warm-starting algorithms \cite{Egger2021warmstartingquantum,beaulieu_max-cut_2021,patti_markov_2021} have been proposed to tackle this issue. Especially, the classical Metropolis-Hastings algorithm has been utilized as a warm-start for VQE to avoid the local minima convergence, due to its provable ergodicity and suitability for unnormalized probability distributions \cite{patti_markov_2021}. Moreover, for an ergodic, discrete-time Markov chain, the number of epochs required to reach a certain threshold of convergence is analytically bounded by
\begin{equation}
    \tau \leq \frac{2}{\Phi^2} \ln\left(\frac{1}{\alpha_\text{MC} \sqrt{\pi^*}} \right),
    \label{eq:mixing_time}
\end{equation}
\noindent where $\alpha_{MC} = |S-\pi|$ is the distance between the Markov chain's sampled distribution $S$ and the true stationary distribution $\pi$, $\pi^*$ is the probability of the least likely (maximum energy) state of $\pi$, and $\Phi$ is the conductance of the Markov process \cite{patti_markov_2021,Montenegro2006}. In other words, only a minimal increase of classical overhead
is required to avoid the local minima convergence for QAOA with no increase in the complexity of quantum circuit and the quantum overhead.

In this paper, we introduce a general method to obtain the Ising Hamiltonians for CCOPs and demonstrate its effectiveness in dealing with the minimum weight vertex cover (MWVC) problem, the minimum vertex cover (MVC) problem and the maximal independent set (MIS) problem as examples. We emphasize that the Ising Hamiltonian for the MWVC problem is obtained first time through our method, and the obtained Ising Hamiltonians for the other two examples are in consistent with that discussed in \cite{lucas_ising_2014}.
Moreover, we focus on the Metropolis-Hastings warm-starting algorithm for $1$-depth QAOA. We first provide the analytic form of the loss function for $1$-depth QAOA applied to combinatorial optimization problems and then modify the proposal distribution used in \cite{patti_markov_2021} to present the warm-starting algorithm discussed in this paper. The presented warm-starting algorithm here is completely classical due to the analytic loss function, implying that the quantum overhead has been extensively reduced with only a minimal increase of classical overhead.

The paper is structured as follows. In Sec.~\ref{sec:HAMILTONIANS}, we introduce the method to obtain the Ising Hamiltonians for CCOPs and demonstrate its effectiveness by using the MWVC, MVC and MIS problems as examples. In Sec.~\ref{sec:QAOA_for_ccop}, we briefly review QAOA and provide the analytic form of the loss function for 1-depth QAOA applied to combinatorial optimization problems. Besides, we also present the Metropolis-Hastings warm-starting algorithm for QAOA. In Sec.~\ref{sec:Experiments}, we numerically illustrate the validity of our approach by applying the QAOA to MWVC cases with different sizes and demonstrate the advantages of the Metropolis-Hastings warm-starting algorithm by using 30 randomly generated MVC cases. We discuss our results and conclude in Sec.~\ref{conclusion}.

\section{THE ISING HAMILTONIANS FOR CCOPs}
\label{sec:HAMILTONIANS}
In this section, we study the Ising Hamiltonians for CCOPs, for which the constraint and the target function can be considered as combinatorial optimization problems $P_a$ and $P_b$, respectively. We assume that $P_a$ and $P_b$ are both minimization problems. This assumption is general since maximization problems can be easily transformed to minimization problems. Thus, a CCOP can be considered to find the optimal solution to the problem $P_b$ in the constrained search space given by the optimal solutions of the problem $P_a$.

The Ising Hamiltonians for combinatorial optimization problems without constraints can be easily obtained. Denote the Ising Hamiltonians $H_a$ and $H_b$ for the problems $P_a$ and $P_b$, respectively. Sequentially, the solution to the CCOP can be encoded to the eigenstate corresponding to the minimum eigenvalue of $H_b$ under the constraint that this eigenstate is the ground state of $H_a$. Here, $H_a$ and $H_b$ can be considered to be positive semidefinite due to the fact that $H_a$ and $H_b$ are both diagonal in the computational basis. Let $v_0\textless v_1 \textless \cdots \cdots \textless v_{n_{a}-1}$ and $w_0\textless w_1 \textless \cdots \cdots \textless w_{n_{b}-1}$ denote the different eigenvalues of $H_a$ and $H_b$, respectively. If $v_0 \textless 0$, one can set $H^{'}_a=H_a-v_0I$. $H^{'}_a$ is obviously positive semidefinite. It can also be considered as the Ising Hamiltonian for the problem $P_a$ since it keeps the sorting position of the corresponding eigenvalues. Similarly, one can set $H^{'}_b$ if $w_0 \textless 0$. Below we always assume that $H_a$ and $H_b$ are both positive semidefinite.

We take the Ising Hamiltonian $H_{prob}$ for the CCOP to be the linear combination of $H_a$ and $H_b$,
\begin{equation}\label{formulation_H_{prob}}
H_{prob} = a H_a + b H_b,
\end{equation}
where $a\in\mathbb{R}$ and $b\in\mathbb{R}$.
Denote $\ket{\mathbf{x}_1^{(i)}}, \ket{\mathbf{x}_2^{(i)}}, \cdots \ket{\mathbf{x}_{l_i}^{(i)}}$ the orthogonal eigenstates corresponding to the eigenvalue $w_i$ of $H_b$. It is easy to see that $\left\{\ket{\mathbf{x}_1^{(0)}}, \cdots \ket{\mathbf{x}_{l_0}^{(0)}}, \cdots, \ket{\mathbf{x}_1^{(n_b-1)}}, \cdots \ket{\mathbf{x}_{l_{n_b-1}}^{(n_b-1)}}\right\}$ also forms the eigenstate space for $H_a$ and $H_{prob}$. Thus, we can define
\begin{equation}\label{define_e}
e_i \equiv {\underset {j}{\operatorname{min}}} \{{\lambda}_j^{(i)}\mid H_a \ket{\mathbf{x}_j^{(i)}} = {\lambda}_j^{(i)} \ket{\mathbf{x}_j^{(i)}}\}.
\end{equation}
For the convenience of illustration, we assume that the solution to the CCOP is unique. Setting $o \equiv {\underset {j}{\operatorname{min}}} \{j\mid e_j=v_0\}$, we have

$Theorem\ \emph{1}.$ Let $H_a$ and $H_b$ be the Ising Hamiltonians for the optimization problems $P_a$ and $P_b$, respectively. Assume that the upper bound $U$ of $w_o-w_i$ and the lower bound $L$ of $e_i-v_0$ (or $e_i-e_o$), $i<o$, can be obtained from the eigenspectrum analysis of $H_a$ and $H_b$. The Ising Hamiltonian $H_{prob}$ for the CCOP has the following form,
\begin{equation}\label{H_p_expression_original}
H_{prob} = aH_a+bH_b,
\end{equation}
where $a>0$, $b>0$ and $a>bU/L$.

$Proof.$ By the definitions of $w_i$, $e_i$ and $o$, we have
\begin{align}
&\lambda_j^{(i)}\ge e_i>v_0,\ \ \ \ if\ \ i<o ~and~ j=1, \cdots, l_i, \label{lambda_inequality}\\
&\lambda_k^{(o)}=v_0, \ \ \ \ \ \ \ \ \ \ \ \text{existing }~k \in (1, \cdots, l_o),\label{target_lambda}\\
&w_i>w_o,\ \ \ \ \ \ \ \ \ \ \ \ \ if\ \ i>o. \label{w_inequality}
\end{align}
The inequality~\eqref{lambda_inequality} indicates that the target state which encodes the solution to the CCOP is not one of the eigenstates corresponding to the eigenvalue $w_i$ ($i<o$) of $H_b$, since they are not the ground states of $H_a$. Thus, it can be seen that the target state is $\ket{\mathbf{x}_k^{(o)}}$ with the help of equality~\eqref{target_lambda} and inequality~\eqref{w_inequality}.

In order for that $H_{prob}$ is the Hamiltonian for the CCOP, $\ket{\mathbf{x}_k^{(o)}}$ needs to be the unique ground state of $H_{prob}$. Since $H_{prob}\ket{\mathbf{x}_k^{(o)}} = (av_0+bw_o)\ket{\mathbf{x}_k^{(o)}}$ and $H_{prob}\ket{\mathbf{x}_j^{(i)}} = (a\lambda_j^{(i)}+bw_i)\ket{\mathbf{x}_j^{(i)}}$, it implies that
\begin{equation}\label{Ising_model_requirement}
a v_0 +b w_o\textless  a \lambda_j^{(i)}+b w_i,
\end{equation}
where $j=1, \cdots, l_i$ for $i=0, \cdots, o-1, o+1, \cdots, n_b-1$
and $j=1, \cdots, k-1, k+1, \cdots, l_i$ for $i=o$.

For $i=o$, the inequality \eqref{Ising_model_requirement} is satisfied for any $a>0$, since $\lambda_j^{(o)}> e_o = v_0$ for $j=1, \cdots, k-1, k+1, \cdots, l_o$.
When $i>o$, from the inequality~\eqref{w_inequality} and that $\lambda_j^{(i)}\ge v_0$ the \eqref{Ising_model_requirement} holds for any $a>0$ and $b>0$. When $i<o$, \eqref{Ising_model_requirement} can be written as
\begin{equation}\label{Ising_model_requirement_1}
a>b \frac{w_o-w_i}{e_i-v_0}
\end{equation}
by using the inequality~\eqref{lambda_inequality}, namely, $\lambda_j^{(i)}\ge e_i$. Therefore, we obtain $a>0$, $b>0$ and $a>bU/L$. $\Box$

We illustrate the effectiveness of our theorem by typical problems below.
\subsection{THE ISING HAMILTONIAN FOR THE MVC PROBLEM}
\label{sec:MVC}
Let $G=(V,E)$ be an undirected graph, where $V$ is a set of vertices and $E$ a set of edges, the edge is covered by a vertex set $S \subseteq  V$ when this edge has at least one of its endpoint in $S$. The MVC problem is to find such a vertex set $S$ with the smallest number of vertices under the constraint that every edge of $E$ must be covered. The constraint of the MVC problem can be considered as the optimization problem $P_a$ that searches for the vertex set with the smallest uncovered edges, and the target function of the MVC problem can be considered as the optimization problem $P_b$ that targets to find the vertex set with smallest vertices. We denote the number of the vertices (edges) by $n$ ($m$).

Let the binary bit $z_i$ denotes the $i$th vertex of $G$. $z_i$ is identified with spin down $-1$ when $z_i$ is included in the vertex set $S$, otherwise, spin up $+1$. Similar to the idea introduced in \cite{thomas_monte_2014}, we can use $C_v(\mathbf{z})=\sum_{i=0}^{n-1}\frac{1}{2}(1-z_i)$ to count the number of vertices in $S$. The corresponding Ising Hamiltonian $H_b$ has the form,
\begin{equation}\label{eq:H_b}
H_b=\sum_{i=0}^{n-1}(1-{\sigma_i}^z)/2,
\end{equation}
where ${\sigma_i}^z$ denotes that the standard Pauli operator $Z$ acts on the $i$th spin.

With respect to the problem $P_a$, the edge $<i,j>$ formed by the vertices $z_i$ and $z_j$ is not covered by $S$ only when $z_i=+1$ and $z_j=+1$. Inspired by the idea in \cite{thomas_monte_2014}, the quantity
\begin{equation}\label{eq:P_a_1}
C_e(\mathbf{z})={\underset {<i,j>}{\operatorname{\sum}}} \left(z_i z_j+z_i+z_j+1\right)/4
\end{equation}
can be used to count the number of edges which are uncovered by the set $S$. Thus, the corresponding Ising Hamiltonian $H_a$ is of the form,
\begin{equation}\label{eq:H_a}
H_a={\underset {<i,j>}{\operatorname{\sum}}} \left({\sigma_i}^z{\sigma_j}^z+{\sigma_i}^z+{\sigma_j}^z+I\right)/4.
\end{equation}

From the derivations in Appendix~\ref{sec:appendix_A}, in this case we have
\begin{equation}\label{AP1}
w_o-w_i=o-i,~~~e_{i}-e_{i+1}\ge 1
\end{equation}
for $i<o$. From $e_{i}-e_{i+1}\ge 1$, we have
\begin{equation}
\begin{aligned}
\begin{split}
&e_{o-1}-e_{o}\ge 1,\\ &e_{o-2}-e_{o}=e_{o-2}-e_{o-1}+e_{o-1}-e_{o}\ge 2,\\ &\cdots,\\ &e_{i}-e_{o} \ge o-i,
 \end{split}
 \end{aligned}
 \end{equation}
where $i<o$. Thus, $H_{prob} = aH_a+bH_b$ can be considered as the Ising Hamiltonian for the MVC problem with $a>0$, $b>0$ and $a>\frac{o-i}{o-i}b$, namely, $a>b>0$.

The Ising Hamiltonian for the MVC problem can be simplified as
\begin{equation}
\begin{aligned}
\begin{split}
H_{prob} = {\underset {<i,j>}{\operatorname{\sum}}} \frac{a}{4}\sigma_i^z \sigma_j^z + \sum_{i=0}^{n-1}\left(-\frac{b}{2}+\frac{a}{4}d_i\right)\sigma_i^z,
\end{split}
\end{aligned}
\end{equation}
where $a>b>0$, $d_i$ denotes the degree of the $i$th vertex, and the global phase item has been discarded.

\subsection{The ISING HAMILTONIAN FOR THE MWVC PROBLEM}
\label{sec:MWVC}
An important generalized version of the MVC problem is the MWVC problem, which targets to find such a vertex cover $S \subseteq V$ of minimum total weight under the same constraint as the MVC problem. Different from the graph discussed in Sec.~\ref{sec:MVC}, the $i$th vertex of $G$ related to the MWVC problem is associated with a weight $\alpha_i$. The MWVC problem is NP-hard and more complicated than the MVC problem \cite{Michael_1979,wang_exact_2019}.

The constraint in the MWVC problem can be considered as the optimization problem $P_a$ discussed in Sec.~\ref{sec:MVC}, and the target function can be considered as the optimization problem $P_b$ that targets to find the vertex set with minimum total weight.
Similar to the discussion in Sec.~\ref{sec:MVC}, the Hamiltonians $H_a$ and $H_b$ for the problems $P_a$ and $P_b$ can be expressed as
\begin{equation}\label{eq:MWVC_H_a_b}
\begin{aligned}
\begin{split}
&H_a={\underset {<i,j>}{\operatorname{\sum}}} \left({\sigma_i}^z{\sigma_j}^z+{\sigma_i}^z+{\sigma_j}^z+I\right)/4,\\
&H_b=\sum_{i=0}^{n-1}\frac{\alpha_i}{2}(1-{\sigma_i}^z).
\end{split}
\end{aligned}
\end{equation}
It is straightforward to verify that
\begin{equation}\label{eq:MWVC_eigenvalues}
\begin{aligned}
\begin{split}
&w_o-w_i\le w_{n_b-1}-w_0=\sum_{i=0}^{n-1}\alpha_i-0=\sum_{i=0}^{n-1}\alpha_i,\\
&e_i-e_o\ge 1
\end{split}
\end{aligned}
\end{equation}
for $i<o$. Therefore, we obtain the Ising Hamiltonian for the MWVC problem,
\begin{equation}\label{eq:MWVC_H_prob}
H_{prob}={\underset {<i,j>}{\operatorname{\sum}}} \frac{a}{4}\sigma_i^z \sigma_j^z+\sum_{i=0}^{n-1}\left(-\frac{b}{2}\alpha_i+\frac{a}{4}d_i\right){\sigma_i}^z,
\end{equation}
where $a>\left(\sum_{i=0}^{n-1}\alpha_i\right)b>0$, and the global phase item has been discarded.

\subsection{THE ISING HAMILTONIAN FOR THE MIS PROBLEM}
\label{sec:MIS}
Given a graph $G = (V,E)$, the MIS problem targets at finding an independent set $S\subseteq V$ with the largest number of vertices, where the independent set refers to the set with no edges between vertices.

The constraint of the MIS problem can be considered as the optimization problem $P_a$ that searches for the vertex set $S$ with smallest edges, and the target function can be regarded as the optimization problem $P_b$ that targets to find the vertex set $\overline{S}$ with smallest vertices, where $\overline{S}$ denotes the vertex set $V-S$. The Hamiltonians $H_a$ and $H_b$ for the problems $P_a$ and $P_b$ can be easily derived,
\begin{equation}\label{eq:MIS_H_a_b}
\begin{aligned}
\begin{split}
&H_a={\underset {<i,j>}{\operatorname{\sum}}} \left({\sigma_i}^z{\sigma_j}^z-{\sigma_i}^z-{\sigma_j}^z+I\right)/4,\\
&H_b=nI-\sum_{i=0}^{n-1}(1-{\sigma_i}^z)/2.
\end{split}
\end{aligned}
\end{equation}
For the MIS problem we still have the conclusions $w_o-w_i=o-i$ and $e_{i}-e_{o}\ge o-i$ for $i<o$. The proof is similar to that given in Appendix~\ref{sec:appendix_A}.

Therefore, the Ising Hamiltonian for the MIS problem can be simplified as
\begin{equation}\label{eq:MIS_H_prob}
\begin{aligned}
\begin{split}
H_{prob}={\underset {<i,j>}{\operatorname{\sum}}} \frac{a}{4}\sigma_i^z \sigma_j^z+\sum_{i=0}^{n-1}\left(\frac{b}{2}-\frac{a}{4}d_i\right){\sigma_i}^z,
\end{split}
\end{aligned}
\end{equation}
where $a>b>0$, and the global phase item has been discarded.

\section{STANDARD QAOA and THE METROPOLIS-HASTINGS WARM-STARTING ALGORITHM}
\label{sec:QAOA_for_ccop}
The longitudinal field Ising Hamiltonian has the form,
\begin{equation}\label{eq:Ising}
H_{prob}(\sigma) = -{\underset {<i,j>}{\operatorname{\sum}}} J_{ij}\sigma_i^z \sigma_j^z - \sum_{i=0}^{n-1}h_i\sigma_i^z,
\end{equation}
where $<i,j>$ stands for the interaction between the $i$th and $j$th spins with
strength $J_{ij}$, and $h_i$ represents the longitudinal magnetic field acting on the $i$th spin.
This Hamiltonian can be used to deal with the general combinatorial optimization problems.

\subsection{STANDARD QAOA}
\label{sec:QAOA}
QAOA aims to find the ground state of the Hamiltonian~\eqref{eq:Ising} by minimizing the loss function
\begin{equation}\label{eq:QAOA_function}
F_p(\vec{\gamma},\vec{\beta})=\langle H_{prob}\rangle = \langle \vec{\gamma}, \vec{\beta}|H_{prob}|\vec{\gamma},\vec{\beta}\rangle,
\end{equation}
where $\vec{\gamma}=\left(\gamma_1, \gamma_2, \cdots, \gamma_p\right)$, $\vec{\beta}=\left(\beta_1, \beta_2, \cdots, \beta_p\right)$ and $|\vec{\gamma}, \vec{\beta}\rangle$ is the $p$-depth QAOA ansatz wave function \cite{farhi_quantum_2014,wurtz_maxcut_2021},
\begin{equation}\label{eq:QAOA_ansatz}
\ket{\vec{\gamma},\vec{\beta}}=e^{-i\beta_pH_x}
e^{-i\gamma_pH_{prob}}(\cdots)e^{-i\beta_1H_x}e^{-i\gamma_1H_{prob}}\ket{+}^{\otimes n},
\end{equation}
with $H_x=\sum_i \sigma_i^x$ and $\ket{+}=\frac{\ket{0}+\ket{1}}{\sqrt{2}}$.

$Theorem\ \emph{2}.$  For the cases of $J_{ij}=J$, the analytic expression of $F_1$ has the form,
\begin{small}
\begin{align}\label{F_1_anlytic}
  &F_1(\gamma_1,\beta_1)\nonumber\\
  &=-\frac{J}{2}{\underset {<i,j>}{\operatorname{\sum}}} \Big \{\sin{(4\beta_1)}\cos{(2h_i\gamma_1)}\sin{(-2J\gamma_1)}\cos^{d_i-1}{(2J\gamma_1)}
\nonumber\\&\qquad \quad+\sin{(4\beta_1)}\cos{(2h_j\gamma_1)}\sin{(-2J\gamma_1)}\cos^{d_j-1}{(2J\gamma_1)}
\nonumber\\&\qquad \quad+\sin^2{(2\beta_1)}\sin{(2h_i\gamma_1)}\sin{(2h_j\gamma_1)}
\nonumber\\&\qquad \qquad \qquad\cos^{d_i+d_j-2f_{ij}-2}{(2J\gamma_1)}\left[1+\cos^{f_{ij}}{(4J\gamma_1)}\right]
\nonumber\\&\qquad \quad+\sin^2{(2\beta_1)}\cos{(2h_i\gamma_1)}\cos{(2h_j\gamma_1)}
\nonumber\\&\qquad \qquad \qquad\cos^{d_i+d_j-2f_{ij}-2}{(2J\gamma_1)}\left[1-\cos^{f_{ij}}{(4J\gamma_1)}\right]\Big\}
\nonumber \\&\qquad ~~~~+\sum_{i=0}^{n-1}h_i\sin{2\beta_1}\sin{(2h_i\gamma_1)}\cos^{d_i}{(2J\gamma_1)},
\end{align}
\end{small}%
where $f_{ij}$ is the number of spins that interact with both the $i$th and the $j$th spins.

The proof of Theorem 2 is given in Appendix~\ref{sec:appendix_B} by using the Pauli Solver algorithm \cite{hadfield_quantum_2018}.
Notice that this Theorem is a special case of the work in \cite{ozaeta_expectation_2021}. However, the proof of Theorem 2 in this paper adopts the Pauli Solver algorithm which only utilizes the commutation relations of the Pauli matrices. The relevant calculation can be easily computed in a classical computer by realizing symbolized multiplication rules for Pauli matrices. The angles $(\vec{\gamma}, \vec{\beta})$ can be restricted to the compact set $[0, 2\pi]^p\times[0,\pi]^p$ when $J_{ij} \in \mathbb{Z}$ and $h_i \in \mathbb{Z}$ \cite{farhi_quantum_2014}.
In principle the analytic form of the loss function for QAOA with any depth can be obtained in a similar way.

\subsection{QAOA WITH THE METROPOLIS-HASTINGS WARM-STARTING ALGORITHM}
\label{sec:Warm-starting}
Inspired by the application of classic Metropolis-Hastings techniques to VQE \cite{patti_markov_2021}, we present the Metropolis-Hastings warm-starting algorithm for QAOA in this section.
The Metropolis-Hastings method has particularly useful advantages on sampling in high-dimensional spaces. Its provable ergodicity guarantees that all samples of the distributions are eventually sampled in a statistically representative way, regardless of which initial angle parameter is chosen \cite{Montenegro2006,Kemp_2003,Maslen_2003}.

The QAOA with Metropolis-Hastings warm-starting algorithm contains two parts: $(i)$ choosing the optimal parameters $({\vec{\gamma}}^*,{\vec{\beta}}^*)$ from Algorithm~\ref{algo_warm-starting} (see below) as the initial parameters for standard QAOA; $(ii)$ completing the optimization with a closing sequence of standard QAOA epochs.

Before presenting the Metropolis-Hastings warm-starting algorithm for QAOA with any depth $p$, we first define the Boltzmann distribution
\begin{align}\label{pro_params}
P(\vec{\gamma}^{\ a},\vec{\beta}^{a})\propto exp\left(-\alpha F_p(\vec{\gamma}^{\ a},\vec{\beta}^{a})\right),
\end{align}
where $\alpha>0$.
From \eqref{pro_params}, we can see that the probability of the parameter sample increases exponentially with the decrease of the corresponding loss function. To avoid a gradient of zero that causes the parameters to stop updating during the Metropolis-Hastings warm-starting process, we select the candidate parameter
$\vec{\gamma\,'}$ and $\vec{\beta'}$ to be of the following form,
\begin{align}
\gamma\,'_{i}&=\gamma_{i}-\eta \partial_{\gamma_i}F_p(\vec{\gamma},\vec{\beta})+\xi \Theta_t,\label{eq:candidate_params1}\\
\beta'_{i}&=\beta_{i}-\eta \partial_{\beta_i}F_p(\vec{\gamma},\vec{\beta})+\xi \Theta_t,\label{eq:candidate_params2}
\end{align}
where $x_{i}$ denotes the $i$th component of $\vec{x}$, and we have added
a normally distributed random noise term $\Theta_r \sim \mathcal{N}(0,1)$
with scale parameter $\xi$ (see Algorithm~\ref{algo_warm-starting} for more detail).
The analytic gradient of the loss function can be easily computed classically by
assuming that the analytic expression for $F_p$ can be computed.
Thus the proposal distribution can be defined as
\begin{align}\label{proposal_distribution}
G({\vec{\gamma}}^{\ '},{\vec{\beta}}^{'}|\vec{\gamma},\vec{\beta})=\prod_{i=1}^{p}g(\gamma_i^{'}|\gamma_{i})g(\beta_i^{'}|\beta_{i}),
\end{align}
where
\begin{align*}
g(\gamma_i^{'}|\gamma_{i})&=\text{pdf}\left[\mathcal{N}(\eta\partial_{\gamma_i}
F_p,\xi^2)\right](\gamma_{i}-\gamma_i^{'}),\\
g(\beta_i^{'}|\beta_{i})&=\text{pdf}\left[\mathcal{N}(\eta\partial_{\beta_i}
F_p,\xi^2)\right](\beta_{i}-\beta_i^{'}).
\end{align*}
Here, we emphasize that the warm-starting algorithm presented here
is completely classical, as a result of the analytic expression for loss function. Hence,
it is more efficient than the warm-starting  algorithm introduced in \cite{patti_markov_2021}.


To illustrate the reasonability and effectiveness of the proposal distribution \eqref{proposal_distribution}, we note that the Markov chain with this proposal distribution is strongly irreducible, since for all $\vec{\gamma\,'},\vec{\beta'},\vec{\gamma},\vec{\beta}$,
\begin{small}
\begin{align}
&G(\vec{\gamma\,'},\vec{\beta'}|\vec{\gamma},\vec{\beta}) \nonumber\\
&= \prod_{i=1}^{p}\frac{1}{2\pi\xi^2}exp[\frac{-(\gamma_{i}-\gamma\,'_i-\eta\partial_{\gamma_i}
F_p)^2-(\beta_{i}-\beta'_i-\eta\partial_{\beta_i}
F_p)^2}{2\xi^2}]> 0 \nonumber
\end{align}
\end{small}%
is satisfied for proper $\xi$. The proof is similar to that given in \cite{patti_markov_2021}.
Thus, the resulting Markov chain is provably ergodic, implying that the parameters near the global minima can be necessarily sampled after some epoches.
When $F_p$ has analytic expressions, the calculation error of the numerical gradient descent considered in \cite{patti_markov_2021} can be removed, which gives rise to the proposal distribution presented here.

Note that the resulting Markov chain is still provably ergodic and can also effectively sample the parameters near the global minima when the loss function is numerically calculated with small estimation error. Therefore, the proposal distribution \eqref{proposal_distribution} is still suitable for QAOA with any depth even the analytic loss function cannot be computed. When the loss function has analytic expressions, the warm-starting algorithm presented is completely classical, namely, this algorithm does not involve any quantum overhead. Nevertheless, each iteration of the warm-starting algorithm uses the same amount of quantum resources as that used in each iteration of the standard QAOA when the loss function is numerically estimated.

\begin{algorithm}
\caption{The Metropolis-Hastings warm-starting algorithm}\label{algo_warm-starting}
\SetInd{0.9em}{1.3em}
  \SetKwInOut{Input}{Input}
  \SetKwInOut{Output}{Output}
  \Input{$(\vec{\gamma}^{\;0},\vec{\beta}^{0})$: the initial parameters;\\
  $P(\vec{\gamma},\vec{\beta})$: the Boltzmann distribution used in the Metropolis-Hastings process;\\
  $T_{\text{max}}$: the maximum number of Markovian epochs;\\
  $G(\vec{\gamma\,'},\vec{\beta'}|\vec{\gamma},\vec{\beta})$: the proposal distribution in the Metropolis-Hastings process;\\
  $\eta$: the learning rate in gradient descent;\\
  $\xi$: the scale parameter of noise added in gradient descent;\\
  }
  \Output{$(\vec{\gamma}^{*},\vec{\beta}^{*})$: the parameters corresponding to the minimal value of the loss function during the Markovian epochs;}
  \BlankLine
  ~set $t=0$, $\vec{\gamma}^{*}=\vec{\gamma}^{\,0}$, $\vec{\beta}^{*}=\vec{\beta}^{0}$\;
  ~\While{$t<T$}{
  generate a noise term $\Theta_t$ from the normal distribution $\mathcal{N}(0,1)$\;
  \tcp{the $i$th components of proposal candidate parameters $\vec{\gamma\,'}$ and $\vec{\beta'}$}{$\gamma\,'_i=\gamma_{i}^{\,t}-\eta \partial_{\gamma_i}F_p+\xi \Theta_t$\;
  $\beta'_i=\beta_{i}^{t}-\eta \partial_{\beta_i}F_p+\xi \Theta_t$\;}
  compute the accept rate: $A=\min
  \left(1,\frac{P(\vec{\gamma\,'},\vec{\beta'})G(\vec{\gamma}^{\,t},\vec{\beta}^{t}|\vec{\gamma\,'},
  \vec{\beta'})}{P(\vec{\gamma}^{\;t},\vec{\beta}^{t})
  G(\vec{\gamma\,'},\vec{\beta'}|\vec{\gamma}^{\;t},\vec{\beta}^{t})}\right)$\;
  generate a sample $u$ from uniform distribution $U(0,1)$\;
  \eIf{$u\leq A$}{$\gamma_{i}^{\;t+1}=\gamma\,'_i$\;
  $\beta_{i}^{t+1}=\beta'_i$\;
  \If{$P(\vec{\gamma}^{\;t+1},\vec{\beta}^{t+1})>P(\vec{\gamma}^{*},\vec{\beta}^{*})$}
  {$\vec{\gamma}^{*}=\vec{\gamma}^{\;t+1}$, $\vec{\beta}^{*}=\vec{\beta}^{\;t+1}$\;}}{$\vec{\gamma}^{\;t+1}=\vec{\gamma}^{\;t}$\;
  $\vec{\beta}^{t+1}=\vec{\beta}^{t}$\;}
 $t=t+1$\;
 }
\end{algorithm}

\section{Experimental Section}
\label{sec:Experiments}
In this section, the standard QAOA was applied on the MWVC cases to numerically demonstrate the validity and effectiveness of Theorem 1. Besides, the behavior of the 1-depth QAOA was also compared with and without the Metropolis-Hastings warm-starting algorithm by using 30 randomly generated MVC cases.

\subsection{THE PERFORMANCE OF QAOA ON THE MWVC CASES}
\label{sec:analysis_QAOA For MWVC}
The performance of QAOA implemented on the MWVC cases was analyzed with different sizes based on the Ising Hamiltonians obtained through Theorem 1.
The number of the MWVC cases for each size was selected to be 10 and the cases were generated from the graphs which were all randomly drawn from Erd$\ddot{\text{o}}$s-R$\acute{\text{e}}$nyi ensemble with edge probability 0.5. In order to avoid slow convergence caused by overlarge loss function value, we randomly generate numbers from 0 to 3 per size as the weight values of vertices. This process of randomly generating weights was general since the solution to the MWVC problem was the same when the weight values were all multiplied by a positive number.
Here, 20 initial angle parameters were selected, which were randomly generated from $[-\pi, \pi]^p\times[-\pi,\pi]^p$ and the QAOA was executed, respectively, to get the optimal approximate solution. $b=0.5$ and $a=(\sum_{i=0}^{n-1}\alpha_i)b+0.1$ was chosen for the Ising Hamiltonian to the MWVC cases.

\begin{figure}[htbp]
\centering
\includegraphics[width=8.5cm]{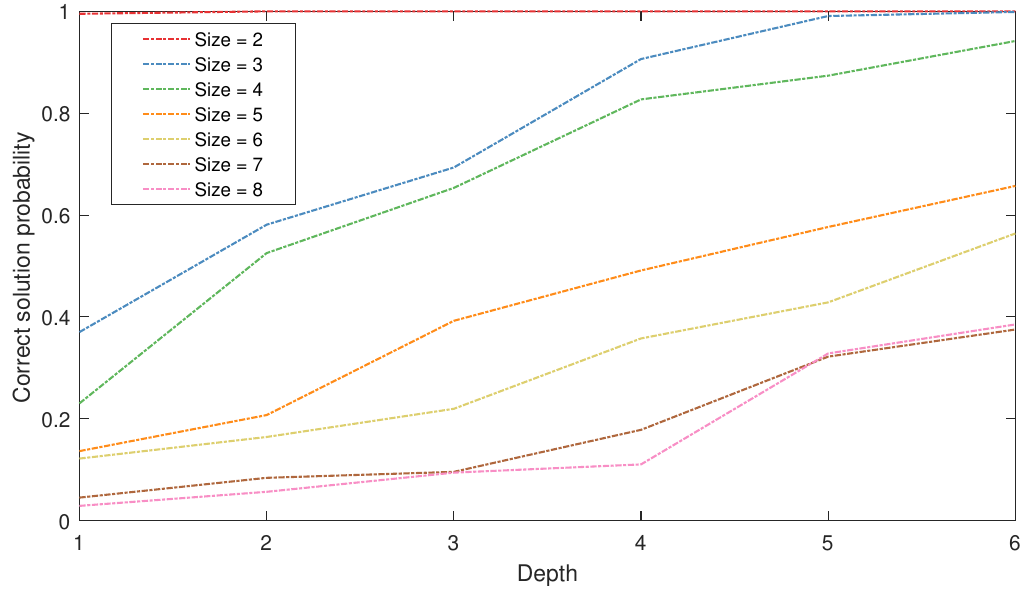}\\
\caption{Correct solution probabilities obtained by implementing standard QAOA on the MWVC cases with different sizes. For the sizes argued here, the correct solution probabilities all raise with the increase of depth.}\label{fig:MWVC_corrent_pro}
\end{figure}

\begin{figure}[htbp]
\centering
\includegraphics[width=8.5cm]{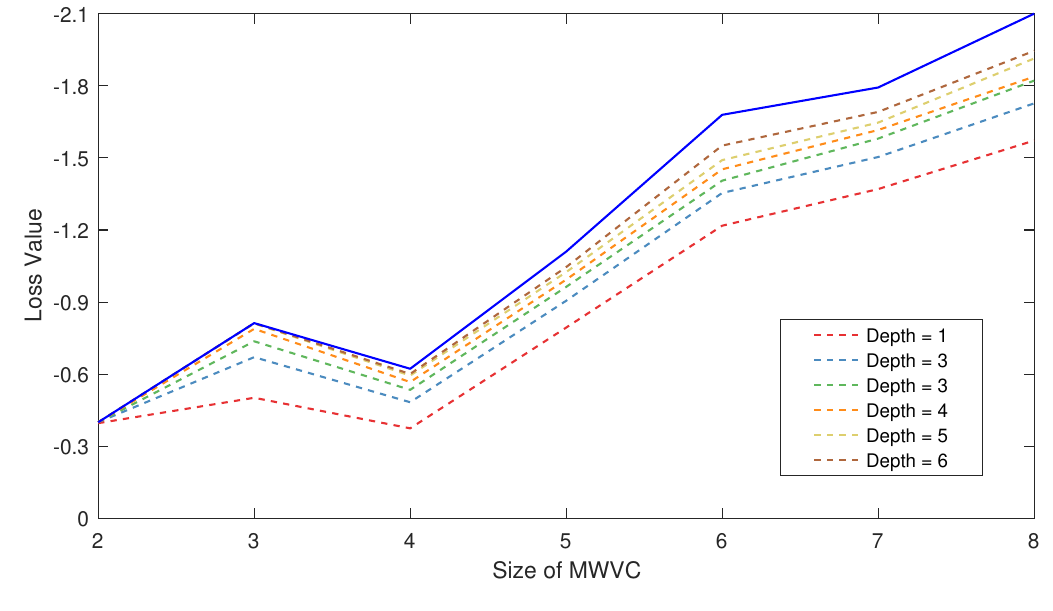}\\
\caption{The loss function values obtained by implementing standard QAOA on the MWVC cases with different sizes. The solid blue line represents the loss function values at the exact solutions. For the sizes argued here, the loss function values all decrease with the increase of depth.}\label{fig:MWVC_loss}
\end{figure}

Brute-force strategy was adopted to obtain the exact solutions for the MWVC cases and the correct solution probability for each size of the MWVC cases was then computed by executing standard QAOA. Here, the correct solution probability for each size refers to the average correct solution probability for 10 MWVC cases with the same size. As shown in Figure~\ref{fig:MWVC_corrent_pro}, the correct solution probabilities all raise with the increase of depth. Correspondingly, the loss function values all decrease with the increase of depth (see Figure~\ref{fig:MWVC_loss}). Moreover, it could be observed that the correct solution probabilities decrease with the increase of the problem size at the same depth which also conforms to the search space characteristics of QAOA.

\subsection{THE PERFORMANCE OF THE METROPOLIS-HASTINGS WARM-START IMPLEMENTED TO QAOA}
\label{sec:analysis_WARM_START}
In this section, the performance of the Metropolis-Hastings warm-starting algorithm was demonstrated by analyzing the average loss function values of QAOA with and without warm-start for 30 MVC cases.

\begin{figure}[htbp]
\centering
\includegraphics[width=8.5cm]{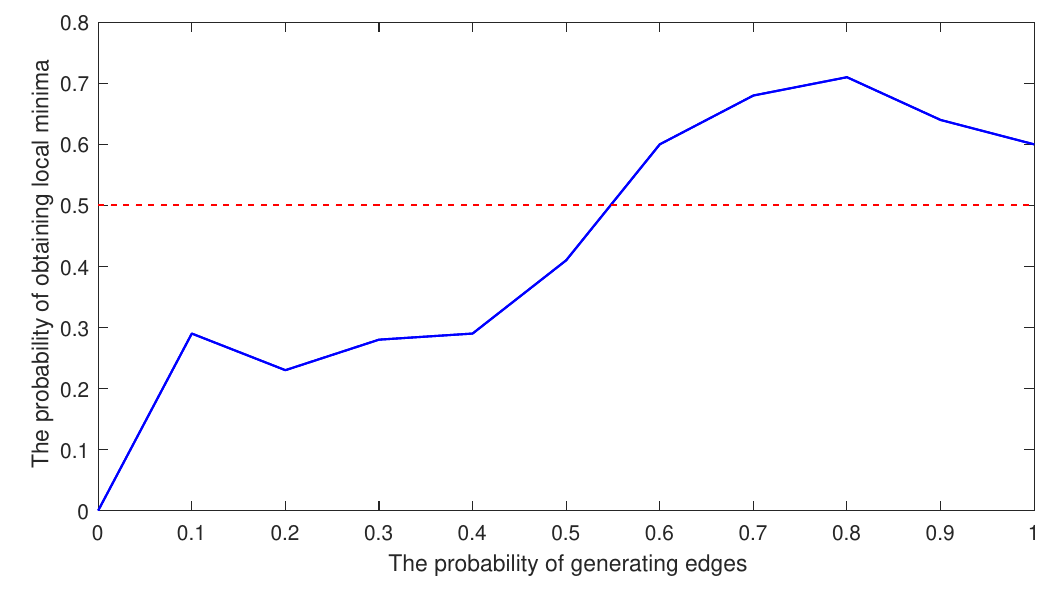}\\
\caption{The probabilities of obtaining the local minima when implementing 1-depth QAOA without warm-start on the MVC cases randomly generated from Erd$\ddot{\text{o}}$s-R$\acute{\text{e}}$nyi ensemble with different edge probabilities. The probability of being trapped in local minima is large than 50\% when the edge probability $\ge$ 0.6.}\label{fig:MVC_QAOA_local_pro}
\end{figure}

To select the cases for which the 1-depth QAOA tends to obtain the local optimal solutions with high probabilities, we first analyze the performance of 1-depth QAOA without warm-start on the graphs randomly generated from Erd$\ddot{\text{o}}$s-R$\acute{\text{e}}$nyi ensemble with different edge probabilities, respectively.  The number of the MVC cases for each edge probability was selected to be 20. From Figure~\ref{fig:MVC_QAOA_local_pro}, it could be observed that the average probability that 1-depth QAOA without warm-start converges to the local minima was larger than 50\% when the graphs were generated from Erd$\ddot{\text{o}}$s-R$\acute{\text{e}}$nyi ensemble with edge probability $\ge$ 0.6. Here, the global minima was considered as the minima obtained by executing 1-depth QAOA with 20 initial angle parameters randomly generated from $[0, 2\pi]^p\times[0,\pi]^p$, respectively. Moreover, it could be easily verified that the number of non-isotropic graphs was small when the ratio of the number of edges to the number of vertices was too large or too small. Thus, we consider the MVC cases generated from 30 graphs with ten vertices which were randomly drawn from the Erd$\ddot{\text{o}}$s-R$\acute{\text{e}}$nyi ensemble with edge probability$=0.6$.

\begin{figure}[htbp]
\centering
\subfigure[]{\includegraphics[width=3in]{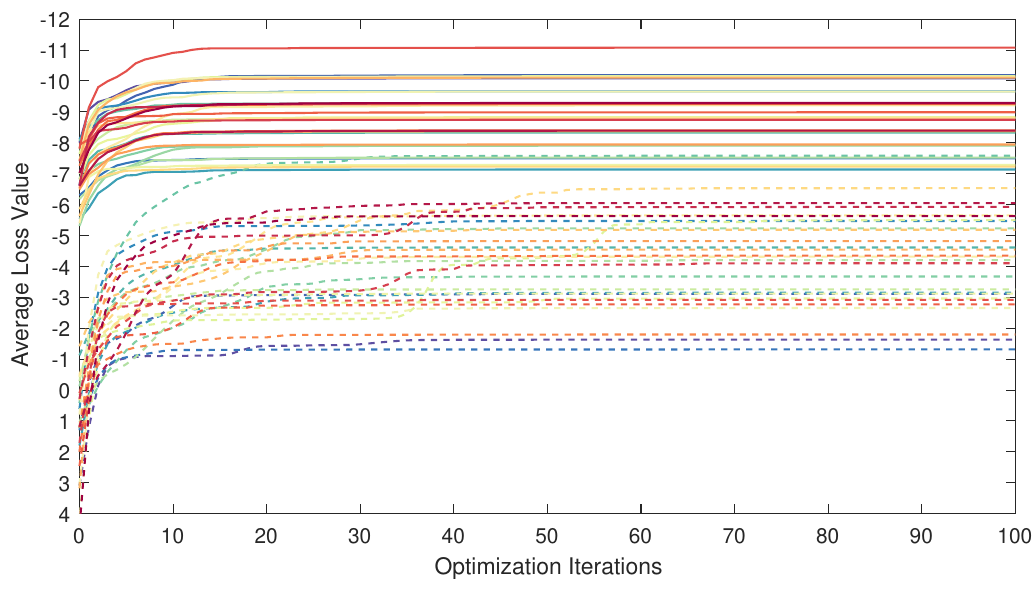}}
\subfigure[]{\includegraphics[width=3in]{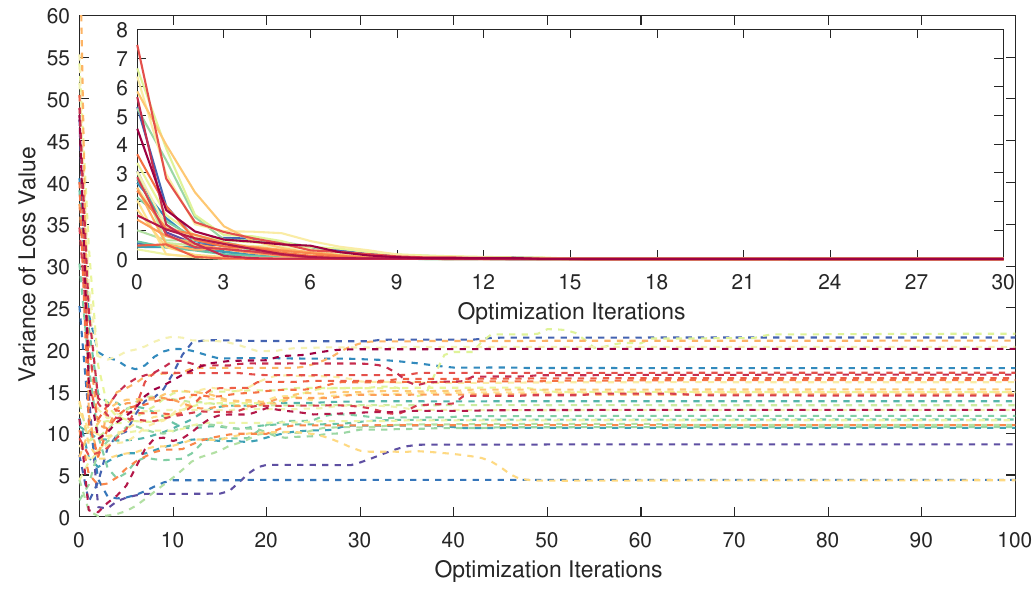}}
\caption{The mean and variance of loss function value of 30 MVC cases for 10 randomly generated initial angle parameters. The solid lines show the numerical results of 1-depth QAOA with warm-start, while the dotted lines represent the numerical results of 1-depth QAOA without warm-start. The numerical results of 1-depth QAOA with and without warm-start for one MVC case adopt the same color. a) The relation between the average loss function value and the optimization iterations for QAOA with and without Metropolis-Hastings warm-start. b) The relation between the variance of loss function value samples and the optimization iterations for QAOA with and without Metropolis-Hastings warm-start.}
\label{fig:MVC_QAOA_samples_loss_and_var}
\end{figure}

In order to show the relation between the initial angle parameters and converging to the global minima, we randomly choose $10$ initial angle parameters from $[0, 2\pi]^p\times[0,\pi]^p$ to execute QAOA with and without Metropolis-Hastings warm-start algorithm, respectively. Here, we set $a=2$ and $b=1$ for the Ising Hamiltonian to the MVC cases. The maximum number of Metropolis-Hastings epochs $T_{\text{max}}$, the parameter $\alpha$ in the Equation~\eqref{pro_params}, the parameter $\xi$ in \eqref{eq:candidate_params1} and \eqref{eq:candidate_params2}, and the learning rate $\eta$ were chosen to be $600$, $0.5$, $0.4$, and $0.1$, respectively. For the 30 MVC cases considered here, the mean of the loss function value samples for the 10 initial angle parameters is shown in Figure~\ref{fig:MVC_QAOA_samples_loss_and_var}(a) and the variance is shown in Figure~\ref{fig:MVC_QAOA_samples_loss_and_var}(b).

From Figure~\ref{fig:MVC_QAOA_samples_loss_and_var}, we see that whether the QAOA without the Metropolis-Hastings warm-start could reach the global minima depends highly on the selection of the initial angle parameters and was irrelevant to the number of optimization iterations. Sequentially, we need to execute the standard QAOA multiple times with different initial angle parameters to get the global optimal results which leads to much quantum overhead.

\begin{figure}[htbp]
\centering
\subfigure[]{\includegraphics[width=3in]{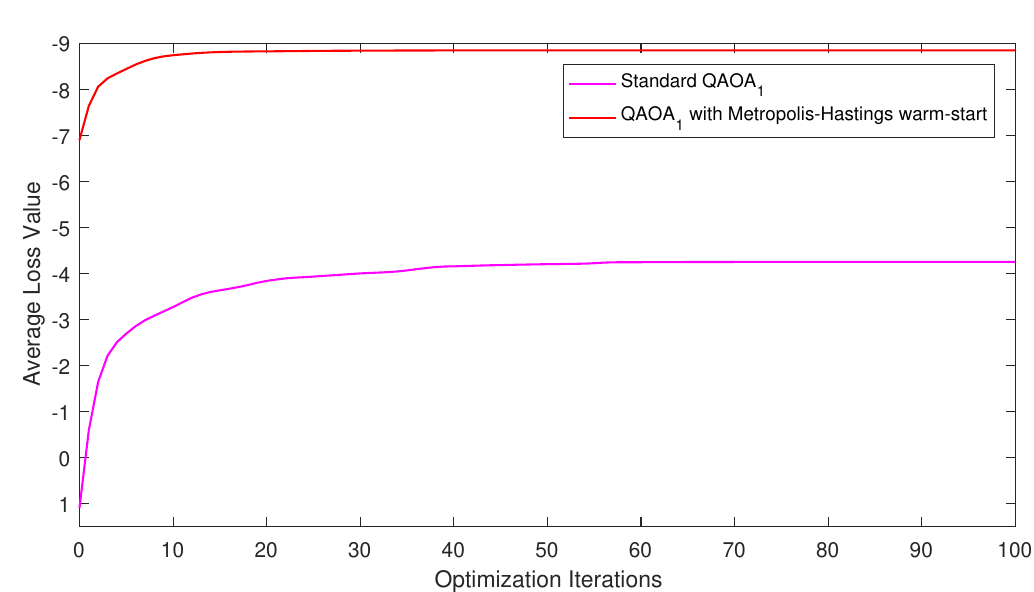}}
\subfigure[]{\includegraphics[width=3in]{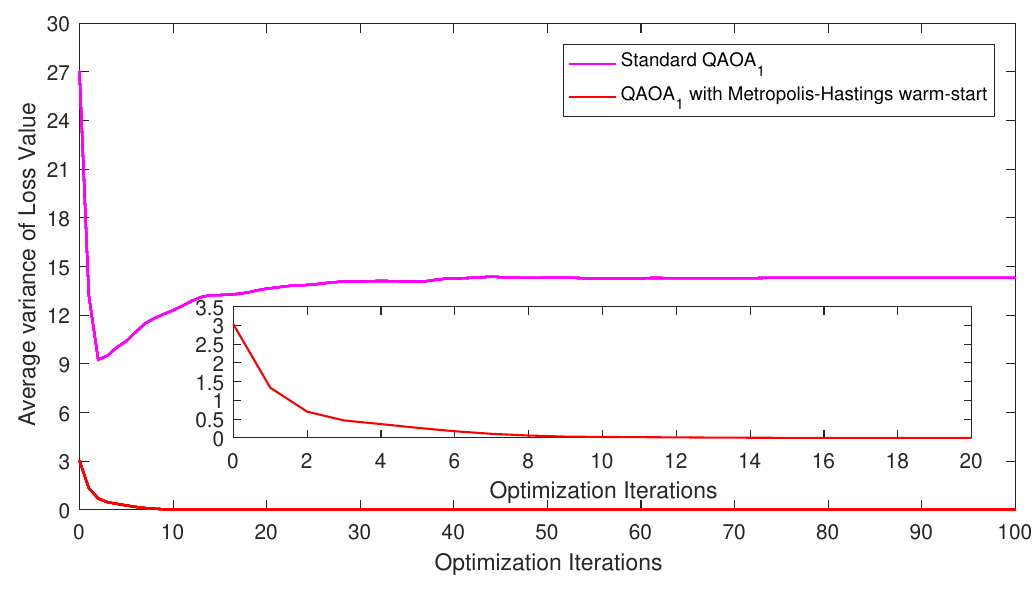}}
\caption{The average mean and variance of loss function value for 30 MVC cases. a) The relation between the average loss function value and the optimization iterations for QAOA with and without Metropolis-Hastings. b)  The relation between the variance of loss function value samples and the optimization iterations for QAOA with and without Metropolis-Hastings warm-start.}
\label{fig:MVC_QAOA_loss_and_var}
\end{figure}

However, for the QAOA with the Metropolis-Hastings warm-start, it could be seen that the influence of initial angel parameters on the converging results was subtle (see Figures~\ref{fig:MVC_QAOA_samples_loss_and_var}(b) and ~\ref{fig:MVC_QAOA_loss_and_var}). In order words, it could always reach the global minima for any randomly generated angle parameters even with low optimization iterations. Therefore, it could dramatically reduce the quantum overhead with the help of Metropolis-Hastings warm-starting algorithm since this warm-starting algorithm was completely classical.


\section{Conclusion}
\label{conclusion}
We have provided a general method to obtain the Ising Hamiltonians for the CCOPs. In order to show the effectiveness of our method, we have derived the Ising Hamiltonians for the MWVC, MVC and MIS problems. The Ising Hamiltonians for the MVC and MIS problems obtained through our method have the same form as that discussed in \cite{lucas_ising_2014}, while the Ising Hamiltonian for the MWVC problem, which is more complicated than the MVC problem, is obtained first time.

We have also applied QAOA to the MWVC cases with different sizes to show the validity and effectiveness of our method numerically. The numerical experiments show that the correct solution probabilities all raise with the increase of depth for the sizes considered in this paper. Correspondingly, the values of loss function all decrease with the increase of the depth. Moreover, it is observed that the correct solution probability decreases with the increase of the problem size for the same depth, which also conforms to the search space characteristics of QAOA.

For the longitudinal field Ising Hamiltonians, which can be used to describe general combinatorial optimization problems as well as other problems such as many-body quantum system problems \cite{chandarana_digitized-counterdiabatic_2022}, we have provided the analytic form of the loss function for 1-depth QAOA. Theoretically, the analytic form of the loss function for QAOA with any depth implemented on longitudinal field Ising Hamiltonians can be obtained by using the Pauli Solver algorithm \cite{hadfield_quantum_2018}. Thus, the analytic form of gradient can be obtained classically, leading to more efficient execution of optimization iterations.

Based on the analytic form of the loss function, we have presented the Metropolis-Hastings warm-starting algorithm for QAOA to obtain the global, rather than local, minima which does not depend on the choice of the initial angle parameters. This warm-starting algorithm is especially efficient for low-depth QAOA as the optimization landscapes become more nonconvex with low-depth.

We emphasize that the warm-starting algorithm discussed here is completely classical, implying that the quantum overhead has been extensively reduced, only with a minimal increase of classical overhead. We have also numerically analyzed the performance of the Metropolis-Hastings warm-starting algorithm for QAOA with depth $p=1$. The numerical experiments show that it can always reach the global minima for any randomly generated angle parameters even with low optimization iterations, while whether the QAOA without the Metropolis-Hastings warm-start can reach global minima depends highly on the selection of initial angle parameters, which can not be improved by increasing the number of optimization iterations.

\begin{acknowledgements} This work was supported by the NSFC (grant nos. 12075159 and 12171044), Beijing Natural Science Foundation (Z190005), and Academician Innovation Platform of Hainan Province.
\end{acknowledgements}

\begin{appendices}
\appendix
\label{appendix}
\section{Proof of (\ref{AP1}) in Sec.~\ref{sec:MVC}}
\label{sec:appendix_A}
\renewcommand{\thesubsection}{A}
It is straightforward to see that the eigenvalues of $H_b$ are given by $0, 1, \cdots, n$. Thus, the conclusion that $w_o-w_i=o-i$ is obvious. Here, we focus on the proof of $e_{i}-e_{i+1}\ge 1$ for $i<o$ below.

Without loss of generality, we assume
\begin{align*}
H_a\ket{\mathbf{x}_p^{(i)}}=e_i\ket{\mathbf{x}_p^{(i)}},
\end{align*}
where $p \in (1, \cdots, l_i)$. Since $\sigma^z \ket{0}=+1\ket{0}$ and $\sigma^z \ket{1}=-1\ket{1}$, the $j$th quantum bit state $\ket{0}$ indicates that the $j$th vertex of $G$ is not included in the vertex set $S$, while the quantum bit state $\ket{1}$ indicates that the $j$th vertex of $G$ is included in $S$. From $e_i>v_0=0$, one has that there exists at least one edge, denoted as $<k,l>$, which is not covered by the vertex set $S$ corresponding to quantum state $\ket{\mathbf{x}_p^{(i)}}$. Thus, the $k$th and $l$th quantum bit states of $\ket{\mathbf{x}_p^{(i)}}$ are both $\ket{0}$.

Next, we set $\ket{\mathbf{y}}=\ket{\mathbf{x}_p^{(i)}}$ except for that the $k$th quantum bit of $\ket{\mathbf{y}}$ is flipped to $\ket{1}$. The vertex set generated by $\ket{\mathbf{y}}$ is denoted as $S_y$ and the number of edges uncovered by $S_y$ is denoted as $\lambda_y$. We can see that the number of vertices included in $S_y$ is $i+1$ and $\lambda_y\le e_i-1$. In other words, we have $H_b\ket{\mathbf{y}}=(i+1)\ket{\mathbf{y}}=w_{i+1}\ket{\mathbf{y}}$ and $H_a\ket{\mathbf{y}}=\lambda_y\ket{\mathbf{y}}$, where $\lambda_y\le e_i-1$. By the definition~\ref{define_e}, we obtained that $e_{i+1}\le \lambda_y\le e_i-1$, namely, $e_i-e_{i+1}\ge1$.

\section{Proof of Theorem 2}\setcounter{equation}{0}
\label{sec:appendix_B}
We prove the Theorem 2 by using the Pauli Solver algorithm introduced in \cite{hadfield_quantum_2018}. It is straightforward to get that
\begin{small}
\begin{align}\label{F_1_express}
&F_1(\gamma_1,\beta_1)\nonumber\\
&=\langle +|^{\otimes{n}}e^{i\gamma_1 H_{prob}}e^{i\beta_1 H_x}( H_{prob}) e^{-i\beta_1 H_x}e^{-i\gamma_1H_{prob}}\ket{+}^{\otimes{n}}
\nonumber\\&=-J{\underset {<u,v>}{\operatorname{\sum}}}\langle +|^{\otimes{n}}e^{i\gamma_1 H_{prob}}e^{i\beta_1 H_x}\left(Z_u Z_v\right)e^{-i\beta_1 H_x}e^{-i\gamma_1 H_{prob}}\ket{+}^{\otimes{n}}
\nonumber\\&\quad-\sum_{u=0}^{n-1}h_u\langle +|^{\otimes{n}}e^{i\gamma_1 H_{prob}}e^{i\beta_1 H_x}\left(Z_u\right) e^{-i\beta_1 H_x}e^{-i\gamma_1 H_{prob}}\ket{+}^{\otimes{n}}
\nonumber\\&=-J{\underset {<u,v>}{\operatorname{\sum}}}\langle Z_u Z_v\rangle-\sum_{u=0}^{n-1}h_u\langle Z_u \rangle,
\end{align}
\end{small}%
where $Z_u=\sigma_u^z$. Let $Q_{uv}=e^{i\gamma_1 H_{prob}}e^{i\beta_1 H_x}Z_u Z_v e^{-i\beta_1 H_x}e^{-i\gamma_1 H_{prob}}$ and  $Q_u=e^{i\gamma_1 H_{prob}}e^{i\beta_1 H_x}Z_u e^{-i\beta_1 H_x}e^{-i\gamma_1 H_{prob}}$.  In order to solve $\langle Z_u Z_v\rangle$ and $\langle Z_u\rangle$ through the Pauli Solver algorithm, we first write
\begin{align}\label{Pauli_expansion_Q}
Q_{uv}=  a_0 I + \sum_{l=0}^{n-1} \sum_{\sigma = X,Y,Z} a_{l\sigma} \sigma_l
+ \sum_{l \neq k } \sum_{\sigma,\lambda = X,Y,Z} a_{lk\sigma\lambda} \sigma_l  \lambda_k + \dots,\nonumber\\
Q_{u}=  b_0 I + \sum_{l=0}^{n-1} \sum_{\sigma = X,Y,Z} b_{l\sigma} \sigma_l
+ \sum_{l \neq k } \sum_{\sigma,\lambda = X,Y,Z} b_{lk\sigma\lambda} \sigma_l  \lambda_k + \dots,
\end{align}
where $a_\alpha \in \mathbb{R}$, $b_\alpha \in \mathbb{R}$ and $X\{Y,Z\}_i$ represents the Pauli matrix $X\{Y,Z\}$ acting on the $i$th spin. Using $\bra{+}I\ket{+}=\bra{+}X\ket{+}=1$ and $\langle +|Y\ket{+}=\langle+|Z\ket{+}=0$, we have
\begin{align}
\langle Z_uZ_v\rangle&= \bra{\gamma_1, \beta_1} Z_u Z_v  \ket{\gamma_1, \beta_1} = \bra{+}^{\otimes n}Q_{uv}\ket{+}^{\otimes n}
 \nonumber\\&= a_0  + \sum_{l=0}^{n-1} a_{lX}
+ \sum_{l \neq k }  a_{lkXX}  + \cdots,
\nonumber\\
 \langle Z_u\rangle &=\bra{\gamma_1, \beta_1} Z_u \ket{\gamma_1, \beta_1} = \bra{+}^{\otimes n}Q_{u}\ket{+}^{\otimes n}
 \nonumber\\&= b_0  + \sum_{l=0}^{n-1} b_{lX}
+ \sum_{l \neq k }  b_{lkXX}  + \cdots.
\end{align}

To prove Theorem \emph{2}, we first introduce the following relations:
\begin{align}
\sigma_u\prod_{<u,k>}e^{iJ\gamma_1Z_u Z_k}&=\prod_{<u,k>}e^{-iJ\gamma_1Z_u Z_k}\sigma_u,\label{lemma_3}\\
\sigma_uZ_ve^{iJ\gamma_1Z_u Z_v}&=e^{-iJ\gamma_1Z_u Z_v}\sigma_uZ_v,\label{lemma_5}\\
\sigma_u\sigma_ve^{iJ\gamma_1Z_u Z_v}&=e^{iJ\gamma_1Z_u Z_v}\sigma_u\sigma_v,\label{lemma_6}\\
\sigma_uZ_v\prod_{<u,l>,l\neq v}e^{iJ\gamma_1Z_u Z_l}&=\prod_{<u,l>,l\neq v}e^{-iJ\gamma_1Z_u Z_l}\sigma_uZ_v,\label{lemma_7}\\
\sigma_uZ_v\prod_{<j,v>,j\neq u}e^{iJ\gamma_1Z_j Z_v}&=\prod_{<j,v>,j\neq u}e^{iJ\gamma_1Z_j Z_v}\sigma_uZ_v,\label{lemma_8}\\
\sigma_u\sigma_v\prod_{<u,l>,l\neq v}e^{iJ\gamma_1Z_u Z_l}&=\prod_{<u,l>,l\neq v}e^{-iJ\gamma_1Z_u Z_l}\sigma_uZ_v,\label{lemma_9}\\
\sigma_u\sigma_v\prod_{<j,v>,j\neq u}e^{iJ\gamma_1Z_j Z_v}&=\prod_{<j,v>,j\neq u}e^{-iJ\gamma_1Z_j Z_v}\sigma_uZ_v,\label{lemma_10}
\end{align}
where $\sigma_u$ denotes the Pauli operators $X$ or $Y$ acting on the $u$th spin.
These relations can be proved straightforwardly by using the properties of
Pauli matrices $X$, $Y$ and $Z$. Here, we prove \eqref{lemma_10} as an example. Since $[X,Z]=-[Z,X],[Y,Z]=-[Z,Y]$, we have
\begin{align}\label{lamma_prove}
\sigma_u\sigma_ve^{iJ\gamma_1Z_j Z_v} &=\sigma_u\sigma_v\left(\cos{(J\gamma_1)}+i\sin{(J\gamma_1)}Z_jZ_v\right)\nonumber\\
&=\left(\cos{(J\gamma_1)}-i\sin{(J\gamma_1)}Z_jZ_v\right)\sigma_u\sigma_v\nonumber\\
&=e^{-iJ\gamma_1Z_j Z_v}\sigma_u\sigma_v,
\end{align}
where $j\neq u$. Thus \eqref{lemma_10} is proved by repeating the process~\eqref{lamma_prove} for different $j$.

Now we prove the Theorem 2 by computing $\langle Z_u\rangle$ and $\langle Z_uZ_v\rangle$.
Since $e^{-i \beta_1 H_x}=\prod_{l=0}^{n-1} e^{-i\beta_1 X_l}$ and $Z_u e^{-i \beta_1 X_u} = e^{i \beta_1 X_u}Z_u$, we have
\begin{align}  \label{eq:mixOpConj_Q_i}
e^{i \beta_1 H_x} Z_u e^{-i \beta_1 H_x} &= e^{2i \beta_1 X_u} Z_u\nonumber\\&=\sin{(2\beta_1)}Y_u+\cos{(2\beta_1)}Z_u.
\end{align}
Since $Z_u$ in the second term commutes with $e^{-i\gamma_1 H_{prob}}$, it does not contribute to $\langle Z_u\rangle$.

For $Y_u$ in the first term in \eqref{eq:mixOpConj_Q_i}, it is straightforward to verify that
\begin{align}\label{H_prob_self_Y_u}
\prod_{l=0}^{n-1}e^{-i\gamma_1 h_lZ_l}\left(Y_u\right)\prod_{l=0}^{n-1}e^{i\gamma_1 h_lZ_l}&=e^{-i\gamma_1 h_uZ_u}\left(Y_u\right)e^{i\gamma_1 h_uZ_u}\nonumber\\
&=e^{-2i\gamma_1 h_uZ_u}\left(Y_u\right)\nonumber\\
&=\sin{(-2h_u\gamma_1)}X_u+\cos{(2h_u\gamma_1)}Y_u.
\end{align}
With the help of \eqref{lemma_3}, we have
\begin{align}
&\prod_{<j,k>}e^{-iJ\gamma_1Z_j Z_k} \left(\sigma_u\right) \prod_{<j,k>}e^{iJ\gamma_1Z_j Z_k}\nonumber\\
&=\prod_{<u,k>}e^{-iJ\gamma_1Z_u Z_k}\left(\sigma_u\right)\prod_{<u,k>}e^{iJ\gamma_1Z_u Z_k}
\nonumber\\
&=\prod_{<u,k>}e^{-2iJ\gamma_1Z_u Z_k}\left(\sigma_u\right).
\end{align}
Since
\begin{small}
\begin{align} \label{H_prob_interact_X_u}
\prod_{<u,k>}e^{-2iJ\gamma_1Z_u Z_k}\left(X_u\right)=\prod_{<u,k>}\left[\cos{(2J\gamma_1)}-i\sin{(2J\gamma_1)}Z_uZ_k\right]\left(X_u\right),
\end{align}
\end{small}%
the only term which contributes to the expectation value $\langle Z_u\rangle$ is $\cos^{d_u}{(2J\gamma_1)}X_u$
in the expanded right hand side.

Similarly, since
\begin{small}
\begin{align}\label{H_prob_interact_Y_u}
\prod_{<u,k>}e^{-2iJ\gamma_1Z_u Z_k}\left(Y_u\right)=\prod_{<u,k>}\left[\cos{(2J\gamma_1)}-i\sin{(2J\gamma_1)}Z_uZ_k\right]\left(Y_u\right),
\end{align}
\end{small}%
the right hand side of \eqref{H_prob_interact_Y_u} does not contribute to $\langle Z_u\rangle$. Thus, we have
\begin{align}\label{H_prob_Y_u}
\bra{+}^{\otimes n} e^{i\gamma_1H_{prob}}Y_ue^{-i\gamma_1H_{prob}}\ket{+}^{\otimes n}=\sin(-2h_u\gamma_1)\cos^{d_u}{(2J\gamma_1)}.
\end{align}

Combining \eqref{H_prob_Y_u} and \eqref{eq:mixOpConj_Q_i}, we have
\begin{align}\label{<Z_u>}
\langle Z_u\rangle=\bra{+}^{\otimes n} e^{i\gamma_1H_{prob}}e^{i\beta_1H_x}Z_ue^{-i\beta_1H_x}e^{-i\gamma_1H_{prob}}\ket{+}^{\otimes n}\nonumber\\=\sin(2\beta_1)\sin(-2h_u\gamma_1)\cos^{d_u}{(2J\gamma_1)}.
\end{align}
Corresponding to \eqref{eq:mixOpConj_Q_i}, we have
\begin{small}
\begin{align}
e^{i \beta_1 H_x} Z_u Z_v e^{-i \beta_1 H_x} &= e^{2i \beta_1 X_u} e^{2i \beta_1 X_v} Z_u Z_v\nonumber
\\&=\sin^2{(2\beta_1)}Y_uY_v+\frac{1}{2}\sin{(4\beta_1)}(Y_uZ_v+Z_uY_v)\nonumber
\\&\quad+\cos^2{(2\beta_1)}Z_uZ_v,\label{eq:mixOpConj_Q_ij}
\end{align}
\end{small}%
by taking into the the fact that $Z_u Z_ve^{-i \beta_1 X_u}e^{-i \beta_1 X_v} = e^{i \beta_1 X_u}e^{i \beta_1 X_v}Z_u Z_v$. The term $Z_uZ_v$ commutes with $e^{(-i\gamma_1H_{prob})}$ and thus does not contribute to $\langle Z_uZ_v\rangle$.

For the term $Y_uY_v$ in \eqref{eq:mixOpConj_Q_ij}, it is straightforward to verify that
\begin{align}
&\prod_{l=0}^{n-1}e^{-i\gamma_1 h_lZ_l}\left(Y_uY_v\right)\prod_{l=0}^{n-1}e^{i\gamma_1 h_lZ_l}\nonumber\\
&=e^{-i\gamma_1 h_uZ_u}e^{-i\gamma_1 h_vZ_v}\left(Y_uY_v\right)e^{i\gamma_1 h_vZ_u}e^{i\gamma_1 h_uZ_v}\nonumber\\
&=e^{-2i\gamma_1 h_uZ_u}e^{-2i\gamma_1 h_vZ_v}\left(Y_uY_v\right)\nonumber\\
&=\sin(2h_u\gamma_1)\sin(2h_v\gamma_1)X_uX_v\nonumber\\
&~~~+\cos(2h_u\gamma_1)\sin(-2h_v\gamma_1)Y_uX_v\nonumber\\
&~~~+\sin(-2h_u\gamma_1)\cos(2h_v\gamma_1)X_uY_v\nonumber\\
&~~~+\cos(2h_u\gamma_1)\cos(2h_v\gamma_1)Y_uY_v,\label{eq:H_prob_self_Y_uY_v}
\end{align}
by using the fact $Y_je^{i\gamma_1h_jZ_j}=e^{-i\gamma_1h_jZ_j}Y_j$.

Similarly, for the terms $Y_uZ_v$ and $Z_uY_v$ in \eqref{eq:mixOpConj_Q_ij}, we have
\begin{align}
&\prod_{l=0}^{n-1}e^{-i\gamma_1 h_lZ_l}\left(Y_uZ_v\right)\prod_{l=0}^{n-1}e^{i\gamma_1 h_lZ_l}\nonumber\\&\qquad=
e^{-i\gamma_1 h_uZ_u}e^{-i\gamma_1 h_vZ_v}\left(Y_uZ_v\right)e^{i\gamma_1 h_uZ_u}e^{i\gamma_1 h_vZ_v}\nonumber\\
&\qquad=e^{-2i\gamma_1 h_uZ_u}\left(Y_uZ_v\right)\nonumber\\
&\qquad=\sin(-2h_u\gamma_1)X_uZ_v+\cos(2h_u\gamma_1)Y_uZ_v,\label{eq:H_prob_self_Y_uZ_v}\\
&\prod_{l=0}^{n-1}e^{-i\gamma_1 h_lZ_l}\left(Z_uY_v\right)\prod_{l=0}^{n-1}e^{i\gamma_1 h_lZ_l}\nonumber\\&\qquad=
e^{-i\gamma_1 h_uZ_u}e^{-i\gamma_1 h_uZ_v}\left(Z_uY_v\right)e^{i\gamma_1 h_uZ_u}e^{i\gamma_1 h_vZ_v}\nonumber\\
&\qquad=e^{-2i\gamma_1 h_vZ_v}\left(Z_uY_v\right)\nonumber\\
&\qquad=\sin(-2h_v\gamma_1)Z_uX_v+\cos(2h_v\gamma_1)Z_uY_v.\label{eq:H_prob_self_Z_uY_v}
\end{align}

With respect to the term $X_uZ_v$ in \eqref{eq:H_prob_self_Y_uZ_v}, by using \eqref{lemma_5}, \eqref{lemma_7} and \eqref{lemma_8} we obtain
\begin{align}  \label{eq:Prob_Q_ij_X_uZ_v}
&\prod_{<j,k>}e^{-iJ\gamma_1Z_j Z_k}\left(X_uZ_v\right)\prod_{<j,k>}e^{iJ\gamma_1Z_j Z_k}\nonumber\\
&=e^{-iJ\gamma_1Z_u Z_v}\prod_{<u,l>,l\neq v}e^{-iJ\gamma_1Z_u Z_l}\prod_{<j,v>,j\neq u}e^{-iJ\gamma_1Z_j Z_v}\left(X_uZ_v\right)\nonumber\\
&\quad\prod_{<j,v>,j\neq u}e^{iJ\gamma_1Z_j Z_v}\prod_{<u,l>,l\neq v}e^{iJ\gamma_1Z_u Z_l}e^{iJ\gamma_1Z_u Z_v}\nonumber\\
&=e^{-2iJ\gamma_1Z_u Z_v}\prod_{<u,l>,l\neq v}e^{-2iJ\gamma_1Z_u Z_l}\left(X_uZ_v\right)\nonumber\\
&=\left[\cos(2J\gamma_1)-i\sin(2J\gamma_1)Z_uZ_v\right]\nonumber\\
&\quad\prod_{<u,l>,l\neq v}\left[\cos(2J\gamma_1)-i\sin(2J\gamma_1)Z_uZ_l\right]\left(X_uZ_v\right),
\end{align}
Expanding the product on the right hand side above gives rise to the sum of tensor products of Pauli operators.
Clearly, no term contributes to the expectation value $\langle Z_uZ_v\rangle$, namely,
\begin{align} \label{eq:H_prob_interact_X_uZ_v}
\bra{+}^{\otimes n}\prod_{<j,k>}e^{-iJ\gamma_1Z_j Z_k}\left(X_uZ_v\right)\prod_{<j,k>}e^{iJ\gamma_1Z_j Z_k}\ket{+}^{\otimes n}=0.
\end{align}
This implies by symmetry that
\begin{align} \label{eq:H_prob_interact_Z_uX_v}
\bra{+}^{\otimes n}\prod_{<j,k>}e^{-iJ\gamma_1Z_j Z_k}\left(Z_uX_v\right)\prod_{<j,k>}e^{iJ\gamma_1Z_j Z_k}\ket{+}^{\otimes n}=0.
\end{align}

Concerning the term $Y_uX_v$ in \eqref{eq:H_prob_self_Y_uY_v}, we have from \eqref{lemma_6}, \eqref{lemma_9} and \eqref{lemma_10},
\begin{align}\label{eq:H_prob_interact_Y_uX_v}
&\prod_{<j,k>}e^{-iJ\gamma_1Z_j Z_k}\left(Y_uX_v\right)\prod_{<j,k>}e^{iJ\gamma_1Z_j Z_k}\nonumber\\
&=e^{-iJ\gamma_1Z_u Z_v}\prod_{<u,l>,l\neq v}e^{-iJ\gamma_1Z_u Z_l}\prod_{<j,v>,j\neq u}e^{-iJ\gamma_1Z_j Z_v}\left(Y_uX_v\right)\nonumber\\
&\quad\prod_{<j,v>,j\neq u}e^{iJ\gamma_1Z_j Z_v}\prod_{<u,l>,l\neq v}e^{iJ\gamma_1Z_u Z_l}e^{iJ\gamma_1Z_u Z_v}\nonumber\\
&=\prod_{<u,l>,l\neq v}e^{-2iJ\gamma_1Z_u Z_l}\prod_{<j,v>,j\neq u}e^{-2iJ\gamma_1Z_j Z_v}\left(Y_uX_v\right)\nonumber\\
&=\prod_{<u,l>,l\neq v}\left[\cos(2J\gamma_1-i\sin(2J\gamma_1))Z_uZ_l\right]\nonumber\\
&\quad\prod_{<j,v>,j\neq u}\left[\cos(2J\gamma_1-i\sin(2J\gamma_1))Z_jZ_v\right]\left(Y_uX_v\right).
\end{align}
Expanding the product on the right hand side above, one has the sum of tensor products of Pauli operators. Thus, we see that no terms contribute to the expectation value $\langle Z_uZ_v\rangle$, namely,
\begin{align} \label{contribute_Y_uX_v}
\bra{+}^{\otimes n}\prod_{<j,k>}e^{-iJ\gamma_1Z_j Z_k}\left(Y_uX_v\right)\prod_{<j,k>}e^{iJ\gamma_1Z_j Z_k}\ket{+}^{\otimes n}=0.
\end{align}
By symmetry, this implies that
 \begin{align} \label{contribute_X_uY_v}
\bra{+}^{\otimes n}\prod_{<j,k>}e^{-iJ\gamma_1Z_j Z_k}\left(X_uY_v\right)\prod_{<j,k>}e^{iJ\gamma_1Z_j Z_k}\ket{+}^{\otimes n}=0.
\end{align}
Therefore, we have
\begin{align}
&\bra{+}^{\otimes n}\prod_{<j,k>}e^{-iJ\gamma_1Z_j Z_k}\left(Y_uZ_v\right)\prod_{<j,k>}e^{iJ\gamma_1Z_j Z_k}\ket{+}^{\otimes n}\nonumber\\
&\qquad=\sin(-2J\gamma_1)\cos^{d_u-1}(2J\gamma_1),\label{contribute_Y_uZ_v}\\
&\bra{+}^{\otimes n}\prod_{<j,k>}e^{-iJ\gamma_1Z_j Z_k}\left(Z_uY_v\right)\prod_{<j,k>}e^{iJ\gamma_1Z_j Z_k}\ket{+}^{\otimes n}\nonumber\\
&\qquad=\sin(-2J\gamma_1)\cos^{d_v-1}(2J\gamma_1),\label{contribute_Z_uY_v}\\
&\bra{+}^{\otimes n}\prod_{<j,k>}e^{-iJ\gamma_1Z_j Z_k}\left(Y_uY_v\right)\prod_{<j,k>}e^{iJ\gamma_1Z_j Z_k}\ket{+}^{\otimes n}\nonumber\\
&\qquad=\frac{1}{2}\cos^{(d_u+d_v-2f-2)}(2J\gamma_1)\left(1-\cos^f(4J\gamma_1)\right).
\end{align}

From \eqref{lemma_6}, \eqref{lemma_9} and \eqref{lemma_10} we have for the term $X_uX_v$,
\begin{align}\label{contribute_X_uX_v}
&\prod_{<j,k>}e^{-iJ\gamma_1Z_j Z_k}\left(X_uX_v\right)\prod_{<j,k>}e^{iJ\gamma_1Z_j Z_k}\nonumber\\
&=e^{-iJ\gamma_1Z_u Z_v}\prod_{<u,l>,l\neq v}e^{-iJ\gamma_1Z_u Z_l}\prod_{<j,v>,j\neq u}e^{-iJ\gamma_1Z_j Z_v}\left(X_uX_v\right)\nonumber\\
&\qquad\prod_{<j,v>,j\neq u}e^{iJ\gamma_1Z_j Z_v}\prod_{<u,l>,l\neq v}e^{iJ\gamma_1Z_u Z_l}e^{iJ\gamma_1Z_u Z_v}\nonumber\\
&=\prod_{<u,l>,l\neq v}e^{-2iJ\gamma_1Z_u Z_l}\prod_{<j,v>,j\neq u}e^{-2iJ\gamma_1Z_j Z_v}\left(X_uX_v\right)\nonumber\\
&=\prod_{<u,l>,l\neq v}\left[\cos(2J\gamma_1)-i\sin(2J\gamma_1)Z_uZ_l\right]\nonumber\\
&\quad\prod_{<j,v>,j\neq u}\left[\cos(2J\gamma_1)-i\sin(2J\gamma_1)Z_jZ_v\right]\left(X_uX_v\right).
\end{align}
In this case, in order for that only operators $I$ or $X$ act on the $u$th and $v$th spins, an even number of spins are required to interact both the $u$th and the $v$th spins. In other words, only the expanded terms
$(Z_uZ_{l_1})*\cdots *(Z_uZ_{l_{2j}})*(Z_{l_1}Z_v)*\cdots *(Z_{l_{2j}}Z_v)*(X_uX_v)$
contribute to the expectation value $\langle Z_uZ_v\rangle$, where $j=0,1,\cdots$. Thus, we have
\begin{small}
\begin{align}
&\bra{+}^{\otimes n}\prod_{<j,k>}e^{-iJ\gamma_1Z_j Z_k}\left(X_uX_v\right)\prod_{<j,k>}e^{iJ\gamma_1Z_j Z_k}\ket{+}^{\otimes n}\nonumber\\
&=\binom{f}{0}\cos^{d_u+d_v-2}(2J\gamma_1)+\binom{f}{2}\cos^{d_u+d_v-2-4}(2J\gamma_1)\sin^{4}(2J\gamma_1)+\cdots\nonumber\\
&=\cos^{d_u+d_v-2-2f}(2J\gamma_1)\sum_{i=0,2,\cdots}^{f}
\binom{f}{i}\left[\cos^2(2J\gamma_1)\right]^{f-i}\left[\sin^2(2J\gamma_1)\right]^{i}
\nonumber\\
&=\frac{1}{2}\cos^{d_u+d_v-2f-2}(2J\gamma_1)\left(1+\cos^f(4J\gamma_1)\right).
\end{align}
\end{small}%
Accounting to that
$$\sum_{i=0,2,\dots}^f \binom{f}{i} a^{f-i} b^{i} = \frac12 ( (a+b)^f + (a-b)^f),$$
we get
\begin{align}
\langle &Z_uZ_v\rangle=\frac{1}{2}\big\{\sin(4\beta_1)\cos(2h_u\gamma_1)\sin(-2J\gamma_1)\cos^{d_u-1}(2J\gamma_1)\nonumber\\
&+\sin(4\beta_1)\cos(2h_v\gamma_1)\sin(-2J\gamma_1)\cos^{d_v-1}(2J\gamma_1)\nonumber\\
&+\sin^2(2\beta_1)\cos(2h_u\gamma_1)\cos(2h_v\gamma_1)\cos^{(d_u+d_v-2f-2)}(2J\gamma_1)\nonumber\\
&~~~\cdot\left(1-\cos^f(4J\gamma_1)\right)\nonumber\\
&+\sin^2(2\beta_1)\sin(2h_u\gamma_1)\sin(2h_v\gamma_1)\cos^{(d_u+d_v-2f-2)}(2J\gamma_1)\nonumber\\
&~~~\cdot\left(1+\cos^f(4J\gamma_1)\right)\big\}.
\end{align}
With all the above discussions, we complete the proof.
\end{appendices}

\begin{thebibliography}{99}
\bibitem{RevModPhys.80.1061} A. Das and B. K. Chakrabarti, Colloquium: Quantum annealing and analog
  quantum computation, \href{https://link-aps-org-s.v.cnu.edu.cn/doi/10.1103/RevModPhys.80.1061}{Rev. Mod. Phys. \textbf{80}, 1061 (2008).}
\bibitem{Farhi2001AQA}  E. Farhi, J. Goldstone, S. Gutmann, J. Lapan, A. Lundgren, and D. Preda, A Quantum
  Adiabatic Evolution Algorithm Applied to Random Instances of an NP-Complete
  Problem, \href{https://www.science.org/doi/abs/10.1126/science.1057726}{Science \textbf{292}, 472 (2001).}
\bibitem{mehta_quantum_2021} V. Mehta, F. Jin, H. De Raedt, and K. Michielsen, Quantum annealing with trigger {Hamiltonians}: {Application} to
  2-satisfiability and nonstoquastic problems, \href{https://link.aps.org/doi/10.1103/PhysRevA.104.032421}{Phys. Rev. A \textbf{104}, 032421 (2021).}
\bibitem{Barends2016DigitizedAQ} R. Barends, A. Shabani, L. Lamata, J. Kelly, A. Mezzacapo, U. L. Heras, R. Babbush, A. G. Fowler, B. Campbell, Y. Chen, \emph {et al.}, Digitized
  adiabatic quantum computing with a superconducting circuit, \href{https://www.nature.com/articles/nature17658}{Nature \textbf{534}, 222 (2016).}
\bibitem{chandarana_digitized-counterdiabatic_2022}  P. Chandarana, N. N. Hegade, K. Paul, F. Albarrán-Arriagada,
E. Solano, A. del Campo, and X. Chen, Digitized-counterdiabatic quantum approximate optimization algorithm, \href{https://link.aps.org/doi/10.1103/PhysRevResearch.4.013141}{Phys. Rev. Research \textbf{4}, 013141 (2022).}
\bibitem{farhi_quantum_2014}  E. Farhi, J. Goldstone, and S. Gutmann, A {Quantum} {Approximate} {Optimization} {Algorithm}, \href{http://arxiv.org/abs/1411.4028}{arXiv:1411.4028 [quant-
ph] (2014).}
\bibitem{wurtz_maxcut_2021} J. Wurtz and P. Love, MaxCut quantum approximate optimization algorithm
  performance guarantees for $p>1$, \href{https://link.aps.org/doi/10.1103/PhysRevA.103.042612}{Phys. Rev.
A \textbf{103}, 042612 (2021).}
\bibitem{crooks_performance_2018} G. E. Crooks, Performance of the {Quantum} {Approximate}
  {Optimization} {Algorithm} on the {Maximum} {Cut} {Problem}, \href{http://arxiv.org/abs/1811.08419}{arXiv:1811.08419 [quant-ph] (2018).}
\bibitem{peruzzo_variational_2014} A. Peruzzo, J. McClean, P. Shadbolt, M.-H. Yung, X.-Q.
Zhou, P. J. Love, A. Aspuru-Guzik, and J. L. O’Brien, A
  variational eigenvalue solver on a photonic quantum processor, \href{http://www.nature.com/articles/ncomms5213}{Nat Commun \textbf{5}, 4213 (2014).}
\bibitem{cerezo_variational_2021} M. Cerezo, A. Arrasmith, R. Babbush, S. C. Benjamin, S. En-do, K. Fujii, J. R. McClean, K. Mitarai, X. Yuan, L. Cincio, \emph {et al.}, Variational {Quantum} {Algorithms}, \href{http://arxiv.org/abs/2012.09265}{Nat Rev Phys \textbf{3}, 625(2021).}
\bibitem{lucas_ising_2014} A. Lucas, Ising formulations of many {NP} problems, \href{https://www.frontiersin.org/articles/10.3389/fphy.2014.00005/full}{Front. Physics \textbf{2}, 5 (2014).}
\bibitem{Preskill2018quantumcomputingin}  J. Preskill, Quantum {C}omputing in the {NISQ} era and
  beyond, \href{https://doi.org/10.22331/q-2018-08-06-79}{Quantum \textbf{2}, 79 (2018).}
\bibitem{farhi_quantum_2000} E. Farhi, J. Goldstone, S. Gutmann,
and M. Sipser, Quantum {Computation} by {Adiabatic} {Evolution}, \href{http://arxiv.org/abs/quant-ph/0001106}{arXiv:quant-ph/0001106 (2000).}
\bibitem{hegade_digitized_2022} N. N. Hegade, X. Chen, and E. Solano, Digitized counterdiabatic quantum optimization, \href{https://journals.aps.org/prresearch/pdf/10.1103/PhysRevResearch.4.L042030}{Phys. Rev. Research \textbf{4}, L042030 (2022).}
\bibitem{hegade_portfolio_2022}   N. N. Hegade, P. Chandarana, K. Paul, X. Chen, F. Albarrán-Arriagada, and E. Solano, Portfolio optimization with digitized counterdiabatic quantum algorithms, \href{https://link.aps.org/doi/10.1103/PhysRevResearch.4.043204}{Phys. Rev. Research
4, 043204 (2022).}
\bibitem{PhysRevA.104.032401} J. Lee, A. B. Magann, H. A. Rabitz, and C. Arenz, Progress toward favorable landscapes in quantum combinatorial optimization, \href{https://link.aps.org/doi/10.1103/PhysRevA.104.032401}{Phys. Rev. A \textbf{104}, 032401 (2021).}
\bibitem{patti_markov_2021} T. L. Patti, O. Shehab, K. Najafi, and S. F. Yelin, Markov {Chain} {Monte}-{Carlo} {Enhanced} {Variational} {Quantum} {Algorithms},\href{http://arxiv.org/abs/2112.02190}{arXiv:2112.02190 [quant-ph] (2021).}
\bibitem{Egger2021warmstartingquantum} D. J. Egger, J. Mare{\v{c}}ek, and S. Woerner, Warm-starting quantum optimization, \href{https://doi.org/10.22331/q-2021-06-17-479}{Quantum \textbf{5}, 479 (2021).}
\bibitem{beaulieu_max-cut_2021} D. Beaulieu and A. Pham, Max-cut {Clustering} {Utilizing} {Warm}-{Start}
  {QAOA} and {IBM} {Runtime}, \href{http://arxiv.org/abs/2108.13464}{arXiv:2108.13464
[quant-ph] (2021).}
\bibitem{Montenegro2006} R. Montenegro and P. Tetali, Mathematical Aspects of Mixing
Times in Markov Chains, \href{https://doi.org/10.1561/0400000003}{ Found. Trends Theor. Comput. Sci. \textbf{1}, 237–354 (2006).}
\bibitem{thomas_monte_2014} N. Thomas, Monte {C}arlo Search for Very Hard {KSAT}
  Realizations for Use in Quantum Annealing, \href{https://arxiv.org/abs/1412.5361v1}{arXiv:1412.5361v1
[cond-mat.stat-mech] (2014).}
\bibitem{Michael_1979} M. R. Garey, A guide to the theory of np-completeness, \href{https://epubs.siam.org/doi/10.1137/1024022}{Computers and intractability (1979).}
\bibitem{wang_exact_2019} L. Wang, C.-M. Li, J. Zhou, B. Jin, and M. Yin, An {Exact}
  {Algorithm} for {Minimum} {Weight} {Vertex} {Cover} {Problem} in {Large}
  {Graphs}, \href{http://arxiv.org/abs/1903.05948}{arXiv:1903.05948 [cs] (2019).}
\bibitem{hadfield_quantum_2018} S. Hadfield, Quantum {Algorithms} for {Scientific}
  {Computing} and {Approximate} {Optimization}, \href{http://arxiv.org/abs/1805.03265}{arXiv:1805.03265v1 [quant-ph] (2018).}
\bibitem{ozaeta_expectation_2021} A. Ozaeta, W. van Dam and P. L. McMahon, Expectation {Values} from the {Single}-{Layer} {Quantum} {Approximate} {Optimization} {Algorithm} on {Ising} {Problems}, \href{https://arxiv.org/abs/2012.03421} {arXiv:2012.03421 [quant-ph] (2021).}
\bibitem{Kemp_2003} F. Kemp, Probability for Statisticians, \href{https://www.semanticscholar.org/paper/Probability-for-Statisticians-Kemp/8a2d460a0102d27f823b9344eb5c4b0cd9d382d4}{Journal of the Royal Statistical Society: Series D (The Statistician) \textbf{52}, 249 (2003).}
\bibitem{Maslen_2003} D. Maslen, The eigenvalues of Kac's master equation, \href{https://doi.org/10.1007/S00209-002-0466-Y}{Mathe-
matische Zeitschrift \textbf{243}, 291 (2003).}
\end{thebibliography}

\end{document}